\documentclass[letterpaper,11pt]{article}
\usepackage{osajnl2,bm,amsfonts,hyperref}

\begin{document}

\title{Wavelength-Diverse Polarization Modulators for Stokes Polarimetry}

\author{Steven Tomczyk, Roberto Casini, Alfred G. de Wijn and Peter G. Nelson}

\affil{High Altitude Observatory,
National Center for Atmospheric Research,\break
P.\ O.\ Box 3000, Boulder, CO 80307-3000, U.S.A.}

\begin{abstract}
Information about the three-dimensional structure of solar magnetic
fields is encoded in the polarized spectra of solar radiation by a
host of physical processes. To extract this information, solar spectra
must be obtained in a variety of magnetically sensitive spectral lines
at high spatial, spectral, and temporal resolution with high
precision. The need to observe many different spectral lines drives
the development of Stokes polarimeters with a high degree of
wavelength diversity. We present a new paradigm for the design of
polarization modulators that operate over a wide wavelength range with
near optimal polarimetric efficiency and are directly applicable to
the next generation of multi-line Stokes polarimeters. These
modulators are not achromatic in the usual sense because their polarimetric
properties vary with wavelength, but they do so in an optimal way. Thus we
refer to these modulators as \textit{polychromatic}. We present here
the theory behind polychromatic modulators, illustrate the concept with
design examples, and present the performance properties of a prototype
polychromatic modulator.
\end{abstract}

\section{Introduction} \label{sec:intro}

Our knowledge of solar magnetism relies heavily on our ability to
detect and interpret the polarization signatures of magnetic fields in
solar spectral lines. Since the identification of sunspots as regions
of strong magnetism by means of the observation of the Zeeman effect
\cite{Ha08}, the number of physical mechanisms applied to the interpretation
of solar spectral line profiles has increased significantly. The Hanle
effect observed in linear polarization has been exploited to diagnose
weak turbulent magnetic fields \cite{St82,TB04,Fa93}. A rich spectrum
of linear polarization observed near the solar limb named the ``second solar
spectrum''  \cite{SK96} has been used to constrain magnetic fields and
scattering physics. Observations of lines like Mn which display
hyperfine splitting have been used to break the degeneracy between
magnetic flux and field for weak magnetic fields \cite{LA02,LA06,AR07}.
Information about plasma kinetics during solar flares is encoded in
the linear polarization of H lines through impact polarization
\cite{Stp07}.

Other important justifications for multi-line observations exist.
Foremost among these is that observations of spectral lines formed at
different heights in the solar atmosphere are needed to constrain the
magnetic field geometry in three dimensions. Secondly, simultaneous
observation of multiple spectral lines and the application of a
line-ratio technique have been shown to greatly enhance the
diagnostic potential of Zeeman-effect observations \cite{St73}. Finally,
improving technologies (e.g. IR detector arrays) have made it possible to
take advantage of the increased sensitivity of the Zeeman effect with
wavelength through the
observation of infrared spectral lines \cite{St87,Ru95}.

It follows that the next generation of Stokes polarimeters
must have the capability to observe the solar atmosphere in a variety of
spectral lines over a wide wavelength range, coupled with the ability
to observe several lines simultaneously. This is reflected in the
design of the recently completed Spectro-Polarimeter for InfraRed and
Optical Regions (SPINOR, \cite{SN06}), installed at the Dunn Solar
Telescope (DST) of the National Solar Observatory on Sacramento Peak
(NSO/SP, Sunspot, NM),
that can observe between 430 and 1600 nm. Post-focus instruments at the
recently completed New Solar Telescope will observe between 400 and
1700 nm \cite{Go03}. The suite of spectro-polarimeters planned for the 4-m
Advanced Technology Solar Telescope will cover the wavelength range
between 380 and 2500 nm \cite{Ke03}. And the proposed science for the planned
4-m European Solar Telescope \cite{Co08} emphasizes multi-wavelength
observations from the UV to the near-IR.

One immediate instrument requirement stemming from this need for
wavelength diversity is that the polarization modulation scheme must
be \textit{efficient} at all wavelengths of interest. Typically,
one attempts to achieve this goal by achromatizing
the polarimetric response of a modulator. This, for instance, is the
rational behind the design of super-achromatic waveplates
\cite{Sam04,Ma08}.
In this paper, instead, we present a new paradigm for the design of
efficient polarization modulators that are not achromatic in the above sense,
since they have polarimetric properties that vary with wavelength.
However, they do so in such a way that they can be operated over a
wide wavelength range with near optimal polarimetric efficiency.
We refer to these modulators as \textit{polychromatic}. Because of
their performance, these are directly applicable to the next generation
of multi-line Stokes polarimeters. In the next section, we summarize the
basic theory behind polarization modulators. Next, we
present example designs
using existing technologies, including a discussion of
the method used to obtain them. Finally, the performance properties
of a prototype polychromatic modulator are presented.

\section{Theory} \label{sec:theory}

Most polarimeters operate by employing retarders and polarizers in a
configuration that encodes polarization information into a modulated
intensity signal. This intensity modulation is measured with a
detector and analyzed to infer the input polarization state. An
extensive treatment of the theoretical operation of Stokes
polarimeters has been given in \cite{Co99,dTC00}. This work draws extensively
from that formalism.

The complete polarization state of an input light beam can be
described by the Stokes vector:
\begin{equation}
\bm{S}\equiv(I,Q,U,V)^T\;,
\end{equation}
where $I$ is the intensity, $Q$ and $U$ are the net linear
polarizations measured in two coordinate frames that
are rotated by $45^\circ$ with respect to each other, and $V$ is
the net circular polarization. Since the Stokes vector contains
four unknown parameters, any polarimeter must make a minimum of
four measurements to determine it.
The polarization properties of a polarimeter in any state can be
conveniently described by a 4$\times$4 Mueller matrix in the usual way.
The first row of
any Mueller matrix captures the transformation of a Stokes vector into
intensity. Assuming that the detector is sensitive to
intensity alone, then only the first row of the polarimeter Mueller matrix is
important.

The modulation of intensity by a $n$-state polarimeter is captured
in the $n\times$4 modulation matrix, $\mathbf{M}$. For each state, the
corresponding row of the modulation matrix (the \textit{modulation vector})
is given by the first row of the Mueller matrix of the polarimeter in
that state. The modulation matrix is generally normalized by the $(1,1)$
element. Note that, for a retarder-based modulator, the first element of
the modulation vector is the same in all modulation states, so after
normalization the first column of the modulation matrix is all made
of 1s. The operation of a polarimeter can then be represented by:
\begin{equation}
\bm{I} = \mathbf{M} \bm{S}\;,
\end{equation}
and the input Stokes vector is obtained by
\begin{equation}
\bm{S} = \mathbf{D} \bm{I}\;,
\end{equation}
where $\mathbf{D}$ is the demodulation matrix. In \cite{dTC00}
it is shown that, for a given modulation matrix, $\mathbf{M}$,
the optimal demodulation matrix is
$\mathbf{D}=(\mathbf{M}^T\mathbf{M})^{-1}\mathbf{M}^T$.
The polarimetric efficiency of a modulation scheme can be
derived from the demodulation matrix itself, and is given by \cite{Co99}:
\begin{equation}
\epsilon_i=\Biggl( n \sum_{j=1}^n D^2_{ij} \Biggr)^{-1/2}\;,
\end{equation}
where the subscript $i$ varies over the four elements of the Stokes
vector. The polarimetric efficiency is important in that it quantifies
the noise propagation through the demodulation process. The noise in
the $i$-th element of the inferred Stokes vector, $\sigma_i$, depends
directly on the efficiency as \cite{dTC00}:
\begin{equation}
\sigma_i=\frac{1}{\sqrt{n}}\frac{\sigma_I}{\epsilon_i}\;,
\end{equation}
where $\sigma_I$ is the uncertainty on the measured intensity, assumed
constant, during the measurement cycle. Equation 5 implies that an
efficient polarization modulator is required to measure a Stokes
vector with high precision. The efficiency on any Stokes parameter can
be between 0 and 1, subject to the two independent
constraints
$\epsilon^2_Q+\epsilon^2_U+\epsilon^2_U\le 1$, and $\epsilon^2_I\le 1$.
Many so-called magnetographs are designed to modulate only Stokes $V$ with high
efficiency. In this paper, we are concerned with modulation
schemes that are balanced, in the sense that the efficiencies of
Stokes $Q$, $U$, and $V$ are equal. In this case, the maximum (i.e., optimal)
modulation efficiency for Stokes $I$ is 1, and for Stokes $Q$, $U$, and $V$,
is $1/\sqrt{3}\approx 57.7\%$.

\begin{table}[htb]
\centering
\begin{tabular}{cccc}
Modulator Type&Retardance&Orientation&Modulation Vector \\
\noalign{\hrule}
\noalign{\vspace{8pt}}
Stepped Retarder&$131.8^\circ$
&$-51.7^\circ$ &$(1,-1/\sqrt{3},+0.375,+0.725)$ \\
&\hbox{''}
&$-15.1^\circ$ &$(1,+1/\sqrt{3},-0.725,+0.375)$ \\
&\hbox{''}
&$+15.1^\circ$ &$(1,+1/\sqrt{3},+0.725,-0.375)$ \\
&\hbox{''}
&$+51.7^\circ$ &$(1,-1/\sqrt{3},-0.375,-0.725)$ \\
\noalign{\vspace{8pt}}
\noalign{\hrule}
\noalign{\vspace{8pt}}
LCVRs (\#1,\#2)
&$(180^\circ,360^\circ)$ &$(0^\circ,45^\circ)$ &$(1,+1,0,0)$ \\
&$(180^\circ,180^\circ)$ &\hbox{''}&$(1,-1,0,0)$ \\
&$(90^\circ,90^\circ)$ &\hbox{''}&$(1,0,+1,0)$ \\
&$(90^\circ,270^\circ)$ &\hbox{''}&$(1,0,-1,0)$ \\
&$(180^\circ,90^\circ)$ &\hbox{''}&$(1,0,0,+1)$ \\
&$(180^\circ,270^\circ)$ &\hbox{''}&$(1,0,0,-1)$ \\
\noalign{\vspace{8pt}}
\noalign{\hrule}
\noalign{\vspace{8pt}}
FLCs (\#1,\#2)
&$(180^\circ,102.2^\circ)$
&$(0^\circ,-18.1^\circ)$ &$(1,+1/\sqrt{3},+1/\sqrt{3},-1/\sqrt{3})$ \\
&\hbox{''}&$(0^\circ,+18.1^\circ)$ &$(1,+1/\sqrt{3},-1/\sqrt{3},+1/\sqrt{3})$ \\
&\hbox{''}&$(45^\circ,+18.1^\circ)$ &$(1,-1/\sqrt{3},+1/\sqrt{3},+1/\sqrt{3})$ \\
&\hbox{''}&$(45^\circ,-18.1^\circ)$ &$(1,-1/\sqrt{3},-1/\sqrt{3},-1/\sqrt{3})$ \\
\noalign{\vspace{8pt}}
\noalign{\hrule}
\end{tabular}
\caption{\label{tab:monochrome}
Examples of optimal modulation schemes for one wavelength,
based on typical retarding devices. The retardances (second column)
are given at the optimization wavelength. The orientation
of the fast axes (third column) are positive counterclockwise, when
looking at the light source. The modulation vector (last column) for
a given $i$-th state corresponds to the $i$-th row of the modulation
matrix. The last example is the modulator solution adopted for the
NSO/DLSP.}
\end{table}

There are several configurations for optimally efficient
polarization modulators at a single wavelength. These may require
different numbers of modulation states to achieve the full measurement
of the Stokes vector. In Table~\ref{tab:monochrome}, we present
examples of optimally
efficient and balanced modulators at a single wavelength, employing
three different modulator technologies. These are a rotating waveplate
of fixed retardation, a pair of liquid crystal variable retarders
(LCVRs) with variable retardation and fixed orientation of the fast
axis (e.g., \cite{To08}), and a pair of ferroelectric liquid crystals
(FLCs) with fixed retardance but variable orientation of the fast
axis. The particular solution presented here for the FLC modulator
is implemented in the Diffraction-Limited Spectro-Polarimeter (DLSP,
\cite{Sa06}), which is also installed at the NSO/SP DST.
An example of a FLC modulator that is optimally
efficient but with higher efficiency for circular polarization is
given in \cite{Ha04}. The LCVR modulator presented here uses 6
modulation states, although it is possible to create a LCVR modulator
that is balanced and optimally efficient at one wavelength
requiring only 4 states. We also note that a single LCVR cannot
modulate all Stokes parameters, regardless of the number
of modulation states.

\begin{figure}[htb]
\centering
\includegraphics[width=88mm]{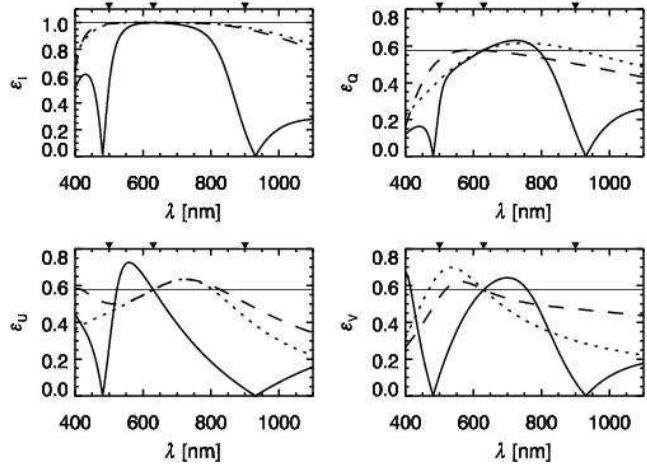}
\caption{\label{fig:flc}
Modulation efficiency curves for the four Stokes parameters between
400 and 1100\,nm for modulators with two FLCs.
\textit{Continuous curve}: optimal and balanced solution at 630\,nm
(indicated by the central arrow) corresponding to the NSO/DLSP
modulator \cite{Sa06}.
\textit{Dotted curve}: polychromatic solution corresponding to a
simple modification of the NSO/DLSP modulator, where the first FLC is
rotated by an additional angle of $67.5^\circ$.
\textit{Dashed curve}: polychromatic solution obtained from the former
solution through the addition of a fixed quartz retarder between the
two FLCs. This solution was optimized between
500 and 900\,nm (indicated by the two outmost arrows).
The horizontal solid lines in the four panels indicate the maximum
theoretical modulation efficiencies that can be achieved simultaneously
for the four Stokes parameters ($1$ for $I$, and $1/\sqrt{3}$ for
$Q$, $U$, and $V$).}
\end{figure}

The retardance of a given device is wavelength dependent so these standard recipes generally work over a limited wavelength
range. One could change the recipe to access other wavelengths but simultaneous spectro-polarimetric observations in different
spectral ranges would not be possible. The chromatic nature of the
example FLC modulator is illustrated in Fig.~\ref{fig:flc} showing the
variation of the efficiency with wavelength. In the plot,
we assume a target wavelength of 630\,nm and that the intrinsic
birefringence of the FLCs is not a function of wavelength (i.e., no
dispersion of the birefringence). We define the region of acceptable
performance of a polarimeter as the region over which the efficiency
is greater than the optimal efficiency divided by $\sqrt{2}$, that is,
$1/\sqrt{6}\approx 40.8\%$.
At this point, the increase in noise is equivalent to a reduction in
the photon flux by a factor of 2, assuming Poisson statistics. By this
definition, we find that the stepped-retarder solution is efficient over a
spectral region of 268\,nm, the LCVR solution over 287\,nm, and the
FLC solution over 128\,nm.

\section{Polychromatic Modulators with Near-Optimal Efficiency} \label{sec:poly}

To create polarimeters that operate over a large wavelength range,
instrument developers have typically tried to select optical materials
in order to make the polarimeter achromatic; that is, to make the
polarimeter modulation matrix as independent of wavelength as
possible. The success of such an approach depends on the choice of
materials and on the desired wavelength coverage. In principle, this
affords an important simplification, which is the possibility of
applying a single demodulation scheme to infer the Stokes vector at
any wavelength. In practice, the application of a single demodulation
scheme can only give a zeroth-order approximation to the true Stokes
vector, because of the unavoidable residual wavelength dependence of
the Mueller matrix of the modulator across the spectral range of
interest. For example, the achromatic modulator of NSO/SPINOR shows a
15\% variation of its retardance properties over its spectral range of
operation (430-1600 nm, \cite{El.PC}). Therefore one still needs to perform a
careful wavelength calibration of the polarimeter in order to infer
the true Stokes vector within the requirements of polarimetric
sensitivity dictated by the science. Previous work \cite{Ga99,Gi03} has also attempted to achromatize modulators by adding combinations of fixed and variable retarders, also
with the goal of minimizing the wavelength dependence of the
resulting Mueller matrix.

However, the preservation of a given form of the polarimeter's
Mueller matrix with wavelength is a very limiting, and arguably
unnecessary, constraint. The fundamental driver in the design of a
polarization modulator for multi-line applications is the
achievement of near-optimal modulation efficiencies in all Stokes
parameters at all wavelengths of interest. The Mueller
matrix of such a modulator can be completely arbitrary, and even
strongly dependent on wavelength. Since calibration with wavelength is
required, such wavelength dependence will not impact the precision of
the polarimetric measurements. In addition, since the demodulation
matrix can be calculated theoretically from the modulator design, it
is straightforward to provide real-time approximations of the
measured Stokes vector at all wavelengths of operation.

In designing the polarimeter for the 1-m Yunnan Solar Telescope
(Yunnan Astronomical Observatory, China) \cite{Xu06} optimized
the efficiency of their polarimeter over a broad wavelength range
by adjusting the orientation of the fast axes of the retarding elements,
and in our opinion created the first polychromatic modulator. However, \cite{Xu06}
did not fully optimize their design by allowing the retardations of the elements to
vary, they did not adopt an optimal demodulation scheme as described in
\cite{dTC00} which could result in lowering the efficiency that is achieved,
and they stressed the importance of minimizing crosstalk which is not necessary
to achieve an optimally efficient polarimeter.

We propose a new paradigm where full-Stokes polarization modulators
are designed to satisfy the constraint of having optimal and
balanced polarimetric efficiency at all wavelengths of interest. This
generally results in polarimeters that have modulation and
demodulation matrices that are strong functions of wavelength. We
refer to these modulators as polychromatic. We have developed a
technique for optimizing polarization modulators using
combinations of fixed and variable retarders. We find that varying
the retardance and orientation of the optical components that comprise
a modulator, and maximizing the polarimetric efficiency for all
wavelengths of interest, results in realizable configurations with a
high degree of wavelength diversity.

We employ an optimization code that systematically searches the
parameter space for a solution that maximizes the Stokes modulation
efficiencies over the desired spectral range. The search is performed
following the strategy of Latin Hypercube Sampling \cite{MK79} of the
parameter space. This significantly improves the convergence speed of
the search for the optimal solution, compared to a direct Monte Carlo
sampling. Although the solutions found through this method are
highly optimized, they may not be unique. Thus it is not possible to
conclude with certainty that they
represent the solutions with the highest possible efficiency for a
given type of modulator. Different optimization schemes based on
gradient methods have also been used with success (e.g., Powell,
Levenberg-Marquardt), although they need realistic first guesses in
order to converge properly.

We illustrate this concept with three different types of polychromatic
modulators for polarimetric applications. For the first type of
modulator, we consider a stack of two FLCs, with the possible addition
of a fixed retarder to broaden the wavelength range. For this type of
modulator, there are four modulation states. The broadening of the
wavelength range of a FLC modulator by the addition of fixed retarders
has been suggested \cite{Ga99} with the intention to minimize off-diagonal
elements of the Mueller matrix, although that study did not result in
efficient modulators over a wide wavelength range.

As a first example, we start with the FLC modulator used by the NSO/DLSP
and presented in Table 1. The DLSP was designed to operate at 630\,nm
only. Figure~\ref{fig:flc} (continuous curve) shows that the DLSP,
as configured, has optimal and balanced efficiency at 630\,nm with
a spectral range of acceptable performance of 128\,nm. We then optimized
the efficiency of the DLSP modulator between 500 and 900\,nm,
by allowing the orientation angles of the FLCs to vary. The resulting
solution (dotted curve) is a simple modification of the NSO/DLSP
configuration (cf.~Table 1), where the first FLC is rotated
such that the new switching angles are $67.5^\circ$ and $112.5^\circ$. With
this new configuration, the modulator still provides optimal and balanced
efficiency at 630\,nm, however the usable range is now twice as large as
before (259\,nm). Next, we added a fixed retarder between the
two FLCs, and allowed the orientations of the FLCs and the retardance
and orientation of the fixed retarder to vary. The resulting solution
(dashed curve) yields a polarimeter with optimal and balanced
efficiency at 630\,nm, but the usable range now spans 553\,nm.
It must be noted that one can
obtain even larger usable ranges with this type of modulator, when
the optimization is extended to all the retarding devices. A clear
example of this (even if still subject to some external constraints)
is the ProMag modulator illustrated in Sect.~\ref{sec:promag}.

\begin{figure}[htb]
\centering
\includegraphics[width=88mm]{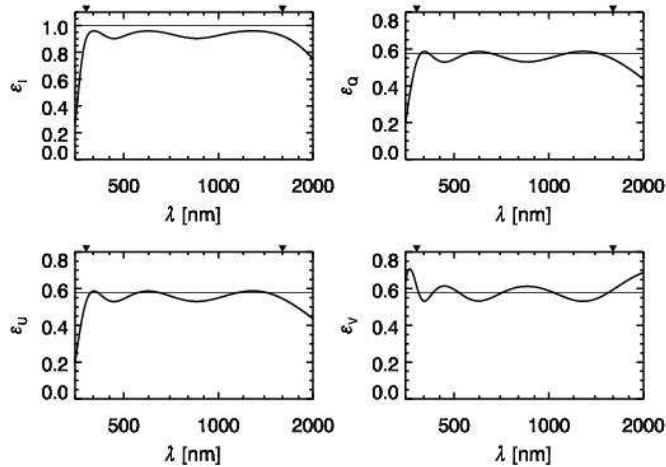}
\caption{\label{fig:ret}
As Fig.~\ref{fig:flc}, but now for a modulator consisting of a stack
of three quartz retarders rotated over 8 discrete steps of 22.5 degrees.
The modulator was optimized between 380 and 1600\,nm (indicated by
the two arrows).}
\end{figure}

The second type of modulator is based on a stack of retarders glued
in a fixed set of relative orientations, that is then rotated as a
whole to the required set of positions to perform the measurement of
the Stokes vector. We consider a stepping modulator, rather than a
continuously rotating one. The adoption of a continuous integration
scheme results in a small reduction of the modulation efficiency. We
refer to \cite{Li87} for a typical estimation of the reduction of
modulation efficiency for continuously rotating modulators. For this
type of rotating modulator, we consider the usual scheme with 8
measurements at positions $k\pi/8$, with $k=0,\ldots,7$. The example
in Fig.~\ref{fig:ret} shows the modulation efficiency of a stack of
three retarders.

\begin{figure}[htb]
\centering
\includegraphics[width=88mm]{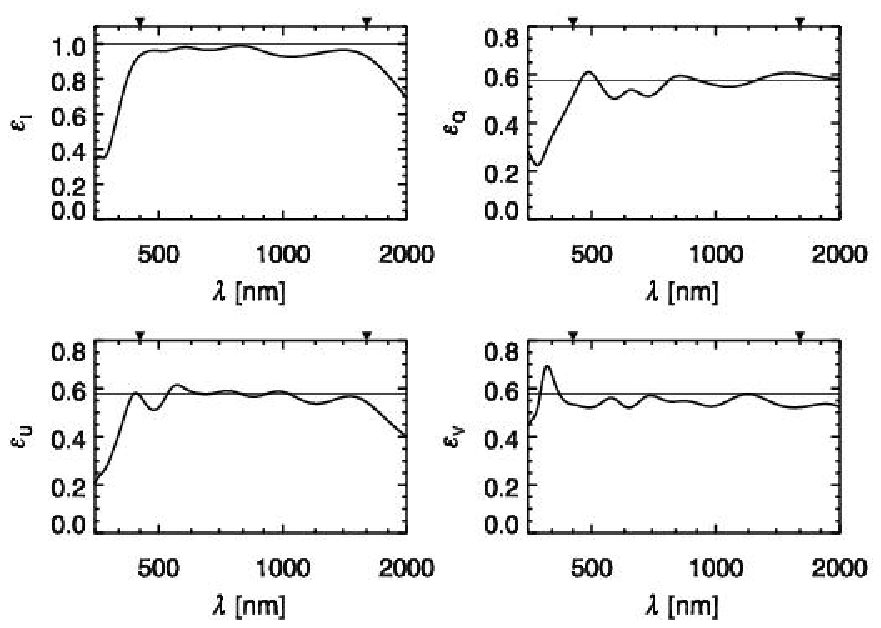}
\caption{\label{fig:lcvr}
As Fig.~\ref{fig:ret}, but now for a modulator consisting of two LCVRs
followed by a fixed quartz retarder, and using a 6-state modulation
scheme. The modulator was optimized between 450 and 1600\,nm
(indicated by the two arrows).}
\end{figure}

The third type is based on a stack of LCVRs followed by a fixed
retarder. The number of states of this kind of modulator is a free
parameter. Figure~\ref{fig:lcvr} considers a modulator with 6 states.
The parameter
space for this kind of modulator has a much higher dimensionality than
for designs based on FLCs or fixed retarders, and therefore multiple
solutions with comparable efficiencies are often found. The example
shown in Fig.~\ref{fig:lcvr} was determined by running the optimization
several
times and selecting the solution with the smallest deviation from optimal
efficiency averaged over the optimization range.

In the solutions shown in Figs.~\ref{fig:ret}--\ref{fig:lcvr}, the
fast axis of the first device is conventionally fixed at $0^\circ$. We verified
that this constraint does not affect the search of an optimally polychromatic
solution for the modulator configuration. For practical purposes, we also
assume that the fixed retarders are made of quartz, while for the LCVRs
we assume no wavelength dispersion of the birefringence.
While this limits the applicability of the recipes given in this paper, the
implementation of realistic birefringence dispersion curves is possible.
Preliminary tests show that this is a mildly limiting factor in the search of
high-efficiency modulator configurations, typically resulting in a narrower
spectral range of optimization for a given number of retarding devices.

Specific science applications of polarimeters may require giving a
higher priority to selected spectral lines and/or Stokes polarization
parameters. For this reason, in our optimization code both the
optimizing wavelengths and the four Stokes parameters can be
attributed different weights in the search of a high-efficiency
modulator configuration. The solutions shown here were derived without
any specific application in mind, and therefore they had equal weights
for all optimizing wavelengths and Stokes parameters. In pratical
situations, tweaking of some of the weights may be necessary to
correct the behavior of the modulation efficiency in a particular
interval of the spectral range. This is not surprising, especially in
the optimization of the modulation efficiency over very large spectral
ranges, since one needs to use a correspondingly large number of
optimizing wavelengths (typically, of the order of 50), and the number
of nearly equivalent solutions increases dramatically because of the
increased number of degrees of freedom in the optimization problem. In
such cases, the use of different weights can help direct the search
towards particular types of nearly equivalent solutions while
excluding others, at the user's discretion.

\section{A Prototype Polychromatic Modulator} \label{sec:promag}

\begin{figure}[htb]
\centering
\includegraphics[width=88mm]{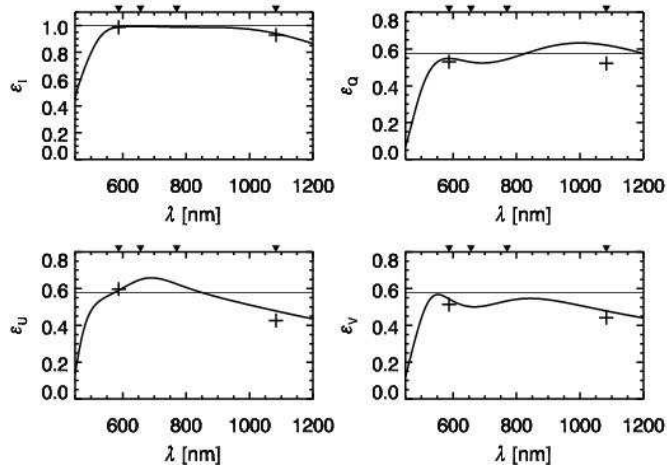}
\caption{\label{fig:promag}
Theoretical efficiency curves for the ProMag modulator, consisting of
two FLCs followed by a quartz retarder. The modulator was optimized at
587.6, 656.3, 769.9, and 1083.0\,nm (indicated by arrows).
The crosses indicate the measured efficiencies at 587.6 and 1083.0\,nm
after deployment of the instrument.}
\end{figure}

As a first application of polychromatic modulators, we re-designed the
polarimeter of the HAO Prominence Magnetometer (ProMag, \cite{El08}). This
instrument was conceived  to observe solar prominences and filaments
in the spectral lines of He I at 587.6 and 1083.0 nm, and also
in H$\alpha$ (656.3 nm). The re-design was prompted by a series of
failed attempts at fabricating the ProMag modulator following the original
design, which was based on a stack of six FLCs \cite{El08}. Thus, the
original design was replaced by a much simpler scheme, utilizing only
two FLCs plus a fixed retarder. The configuration was determined with the
optimization method described above and constrained to use one of the FLCs
from the original design. Retardances of the second FLC and the fixed quartz retarder
were specified according to the optimized solution, and the
retardances of these devices were measured in the laboratory. Since
the measured retardances differed somewhat from the specified ones, a
second optimization was performed varying only the orientations of the
acquired devices using their measured retardances. The modulator was
then constructed according to the design resulting from the second
optimization. We estimate that the elements were oriented to the design values
to within approximately $0.3^\circ$.
Figure~\ref{fig:promag} shows the predicted efficiencies
for the Stokes
parameters resulting from the optimization. Actual efficiencies at
587.6 and 1083.0 nm were measured after deployment of the instrument
at the Evans Solar Facility of the NSO/SP, and they are in
good agreement
with the theory (see crosses in Fig.~\ref{fig:promag}).
The discrepancy observed
at 1083.0 nm is likely caused by a less precise determination of the
retardances of the modulator optics at that wavelength, possibly due
to an undetected leak in the interference filter used during the
measurement in the laboratory, combined with the different
spectral responses of the ProMag Alpha-NIR camera and the photo-diode
used in the lab measurement.

\begin{figure}[htb]
\centering
\includegraphics[width=88mm]{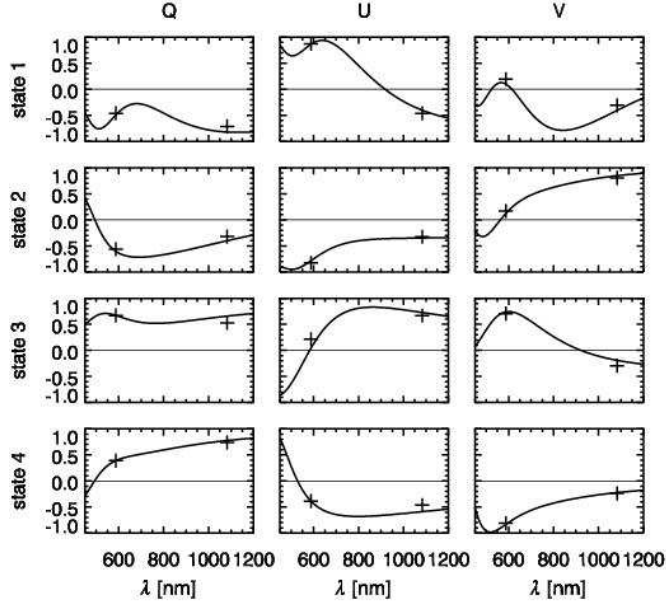}
\caption{\label{fig:promag_matrix}
Theoretical modulation matrix of ProMag. The crosses indicate the
measured modulation amplitudes at 587.6 and 1083.0\,nm. The first
column of the matrix is identically equal to 1 and is not shown.}
\end{figure}

The theoretical variation of the modulation matrix, $\mathbf{M}$, with
wavelength is shown in Fig.~\ref{fig:promag_matrix}. The first column
is unity and therefore is
omitted. Measured modulation amplitudes at 587.6 and 1083.0 nm are
also shown to be in close agreement with the theory.

\section{Conclusions}

In this paper we presented a new paradigm for the design of polarization
modulators that responds to the needs of multi-line spectro-polarimetry.
We argued that the effort of achromatizing the response matrix of a
polarimeter -- that is, trying to make the Mueller matrix as little
dependent on wavelength as possible -- is both a very limiting and
unnecessary constraint. The only requirement that should be
imposed on a polarimeter, in order to reach a given target of
polarimetric sensitivity, is that its modulation efficiency be
sufficiently high at all wavelengths of interest. As these wavelengths
may be distributed over very large spectral ranges (of the order
of 1000\,nm or larger), and because of the wavelength dependence
of the optical properties of retarding devices, the design of
optimally efficient polarization modulators from first principles is a
particularly arduous, if not impossible, task. For this reason, we have
developed an improved Monte Carlo search technique in order to explore
the parameter space of a polarimeter, and identify optimally efficient
solutions compatible with the particular type of modulator. We
illustrated the power of this method by improving the modulation
efficiency of existing polarimetric instruments, as well
as by designing new modulators with near optimal and balanced
efficiency over very large spectral ranges. Finally we have
demonstrated the applicability of this new paradigm by showing the
measured performance of a prototype modulator, which was designed and
optimized through the method presented in this paper.

\section*{Acknowledgments}
We thank D.~Elmore for helpful
discussions during the early stages of this work, S.~Sewell for
helpful comments on the manuscript and Frans Snik for pointing us to
the Yunnan polarimeter paper. The National Center for Atmospheric Research
is sponsored by the National Science Foundation.

\end{document}